\documentclass[structabstract]{aa}  
\usepackage{siunitx}	

\usepackage{graphicx}
\usepackage{txfonts}
\usepackage{amsfonts}
\usepackage{subfigure}

\usepackage{amssymb,amsfonts}
\usepackage{natbib}
\bibpunct{(}{)}{;}{a}{}{,}
\begin{document}
\title{Integrated optics prototype beam combiner for long baseline interferometry in the L and M bands}
\subtitle{} 

   \author{J. Tepper
          \inst{1}\fnmsep\thanks{send correspondence to tepper@ph1.uni-koeln.de}
          \and
          L. Labadie\inst{1}
          \and
		  R. Diener\inst{2}
          \and
          S. Minardi\inst{2,3}
          \and
          J.-U. Pott\inst{4}
          \and
          R. Thomson\inst{5}
          \and
          S. Nolte\inst{2}
          }

\institute{1. Physikalisches Institut, Universit\"at zu K\"oln, Z\"ulpicher Strasse 77, 50937 K\"oln, Germany
\and Institut f\"ur Angewandte Physik, Friedrich-Schiller-Universit\"at Jena, Albert-Einstein-Strasse 15, 07745 Jena, Germany
\and Leibniz-Institut f\"ur Astrophysik Potsdam, An der Sternwarte 16, 14482 Potsdam, Germany
\and Max Planck Institut f\"ur Astronomie, K\"onigstuhl 17, 69117 Heidelberg, Germany
\and Scottish Universities Physics Alliance (SUPA), Institute of Photonics and Quantum Sciences (IPaQS), Heriot Watt University, Riccarton Campus, Edinburgh, EH14 4AS }

\date{Received November 25th, 2016; accepted March 13th, 2017}
  \abstract{Optical long baseline interferometry is a unique way to study astronomical objects at milli-arcsecond resolutions not attainable with current single-dish telescopes. Yet, the significance of its scientfic return strongly depends on a dense coverage of the uv-plane and a highly stable transfer function of the interferometric instrument. In the last few years, integrated optics (IO) beam combiners have facilitated the emergence of 4-telescope interferometers such as PIONIER or GRAVITY, boosting the imaging capabilities of the VLTI. However, the spectral range beyond \SI{2.2}{\micro \meter} is not ideally covered by the conventional silica based IO. Here, we consider new laser-written IO prototypes made of gallium lanthanum sulfide (GLS) glass, a material that permits access to the mid-infrared spectral regime.} 
    {Our goal is to conduct a full characterization of our mid-IR IO two-telescope coupler in order to measure the performance levels directly relevant for long-baseline interferometry. We focus in particular on the exploitation of the L and M astronomical bands. } 
  { 
  We use a dedicated Michelson-interferometer setup to perform Fourier transform spectroscopy on the coupler and measure its broadband interferometric performance. We also analyze the polarization properties of the coupler, the differential dispersion and phase degradation, as well as the modal behavior and the total throughput.}
   {We measure broadband interferometric contrasts of 94.9\% and 92.1\% for unpolarized light in the L and M bands. Spectrally integrated splitting ratios are close to 50\%, but show chromatic dependence over the considered bandwidths. Additionally, the phase variation due to the combiner is measured and does not exceed 0.04\,rad and 0.07\,rad across the L and M band, respectively.
   	 The total throughput of the coupler including Fresnel and injection losses from free-space is 25.4\%. 
   Furthermore, differential birefringence is low ($<$0.2\,rad), in line with the high contrasts reported for unpolarized light.} 
  {The laser-written IO GLS prototype combiners prove to be a reliable technological solution with promising performance for mid-infrared long-baseline interferometry. In the next steps, we will consider more advanced optical functions, as well as a fiber-fed input, and we will revise the optical design parameters in order to further enhance the total throughput and achromatic behavior. }

\authorrunning{J. Tepper}
\titlerunning{Integrated optics for L and M band interferometry}

\maketitle
%

\section{Introduction}\label{Intro}
Aperture synthesis imaging is a major ambition of the optical/IR interferometry community for the next decades and it will remain the only route to reach a level of angular resolution equivalent to that of a diffraction-limited telescope of a few hundred meters’ aperture. In aperture synthesis imaging, the high fidelity of the reconstructed images needed to observe objects with complex morphologies critically depends on our ability to deliver observations with a dense uv coverage \citep{Soulez2016}. Over the last five to ten years, a major incentive has been given to improving this technique at the VLTI and at the CHARA Array. Recently, \cite{Kluska2016} has exploited the four-telescope imaging capabilities of the PIONIER instrument \citep{Pionier} in the H band to evidence a time-variable asymmetry in the close environment of MWC158 whose origin has not yet been determined. Using the same instrument, \cite{Hillen2016} have obtained a direct view of the dust sublimation front in the circumbinary disk of the post-AGB system IRAS 08544-4431. Using the improved uv coverage of the six-telescope MIRC beam combiner operating in the H band, \cite{Roettenbacher2016} were able to map the surface of zeta Andromeda with a 0.5 mas resolution and investigated the surface distribution of starspots to reveal the absence of a solar dynamo mechanism. These recent results highlight the unprecedented potential of optical/infrared interferometric imaging, which has now gone through significant improvement.\newline
\indent The core subsystem of an infrared imaging interferometer is the beam combiner, where the beams of the individual sub apertures coherently interfere. While various designs for beam combination have been explored since the early times of long-baseline interferometry, instrumental solutions based on integrated optics (IO) are now considered serious reliable alternatives to bulk optics designs. Their "on-chip" compact design results in a simpler optical subsystem in comparison to instruments like AMBER \citep{Amber} or MATISSE \citep{Matisse}. 
Moreover, thanks to the single-mode properties of the component, the IO beam combiner delivers a much more stable instrumental transfer function, which is key to the measurement of high accuracy interferometric visibilities. 
After about 15 years of R\&D activities, such concepts have become well-established solutions in the near-IR and are currently implemented in the community instruments GRAVITY \citep{Gravity} and PIONIER \citep{Benisty2009}, in the K and H bands, respectively.\newline
\indent The recognized importance of the mid-infrared spectral range for the study of exoplanetary systems and AGNs at high angular resolution motivates the extension towards longer wavelengths of integrated optics solutions. Due to the intrinsic absorption beyond \SI{3}{\micro \meter} of silica, a specific technological platform for the mid-infrared is needed. The three main technological platforms that have been explored for mid-infrared IO technologies are based on ion exchange/diffusion, chemical etching/lithography, and ultrafast laser writing. \newline
\indent Ion diffusion in lithium niobate glass has demonstrated the feasibility of active IO beam combiners in the 3.2\,-\,\SI{3.8}{\micro \meter} range \citep{Hsiao2009,Heidmann2012}. However, broadband operation by \cite{Martin2014} has evidenced large chromatic dispersion and low-confinement of the modes, which resulted in propagation losses as high as 16 dB/cm.\newline
\indent Etching and lithography techniques have been tested with chalcogenide glass, which raised interest due to its potentially wide mid-infrared transparency from \SI{1}{\micro \meter} to \SI{20}{\micro \meter}. Using this platform, simple rib channel waveguides have been manufactured showing average propagation losses of 0.5-1 dB/cm in the 3\,-\,\SI{6}{\micro \meter} range \citep{Ma2013} and 6 dB/cm in the 2\,-\,\SI{20}{\micro \meter} range \citep{Vigreux2015}. Recently, \cite{KenchingtonGoldsmith2016} used this platform to manufacture a multimode interference coupler (MMI) in chalcogenide glass for nulling interferometry. 
\newline
\indent To date, the technique of direct laser writing has been a successful approach for the manufacture of two-telescope and three-telescope mid-infrared IO beam combiners. Using fluorozirconate ZBLAN glass transparent from \SI{0.2}{\micro \meter} to \SI{5}{\micro \meter}, single-mode channel waveguides with 0.3\,dB/cm losses and evanescent couplers over the spectral range of 3.75\,-\,\SI{4.2}{\micro \meter} were laser-inscribed by \cite{Gross2015}. The laser inscription technique has also been used to manufacture proof-of-concept IO combiners in chalcogenide glass, with reported losses on the order of ~\SI{1}{dB/cm} \citep{Labadie2011,Rodenas2012,Arriola2014}.\newline
\indent In all the cases mentioned above, neither high interferometric contrasts nor a detailed investigation of the differential birefringence and dispersion were reported. These are essential quantities used to assess the potential of IO devices for long-baseline interferometry.
In this paper, we report for the first time to our knowledge a complete performance characterization in the L (3.1\,-\,\SI{3.6}{\micro \meter}) and M bands (4.5\,-\,\SI{4.9}{\micro \meter}) of new 2$\times$2 directional couplers manufactured by laser inscription in a gallium lanthanum sulfide (GLS) chalcogenide glass. This experimental work tests in detail the potential of new IO combiners in the immediate perspective of astronomical applications. The paper is structured as follows. Section 2 presents the design adopted for the integrated optics component and briefly describes the ULI fabrication process. Section 3 describes the laboratory setup and the measurement procedure. Our results are detailed in Sect. 4 and deal with the spectral splitting ratio and modal behavior, the throughput characteristics, the polarization properties of the coupler, and the monochromatic and broadband interferometric performance revealing the impact of the beam combiner on the phase curvature across the L and M bands.
\section{Properties of the integrated optics combiner}
\subsection{Ultrafast laser writing for interferometry}
The 2$\times$2 integrated optics couplers are manufactured using the technique of ultrafast laser inscription (ULI) \citep{Glezer1996,Davis1996,Nolte2003, Thomson2009} that exploits the large peak intensity (up to $10^{12}$\,W/cm$^2$) of a focused femtosecond laser to induce a structural change in the glass sample. This can result in a localized refractive index modification confined in the region where the femtosecond laser is focused. 
Physically, intense light pulses can transfer a substantial fraction of their energy in transparent dielectric media by means of multi-photon ionization followed by avalanche ionization, which could trigger localized structural glass modifications from a chain of chemical and/or thermodynamic resettlements of the glass network. 
Depending on the chosen substrate material, either a local increase or decrease in the bulk refractive index can be observed. By translating  the irradiated sample under the focused laser beam, it is possible to create a pattern of waveguides, with an index difference between the core and the cladding depending basically on the laser power and the duration of irradiation. Recently, this technique has attracted some attention in the field of astronomical instrumentation. Further detail on the technique is available from the review of \cite{Gross2015b}. 
 \begin{figure}[t]
  \centering
  	 \includegraphics[width=0.8\columnwidth]{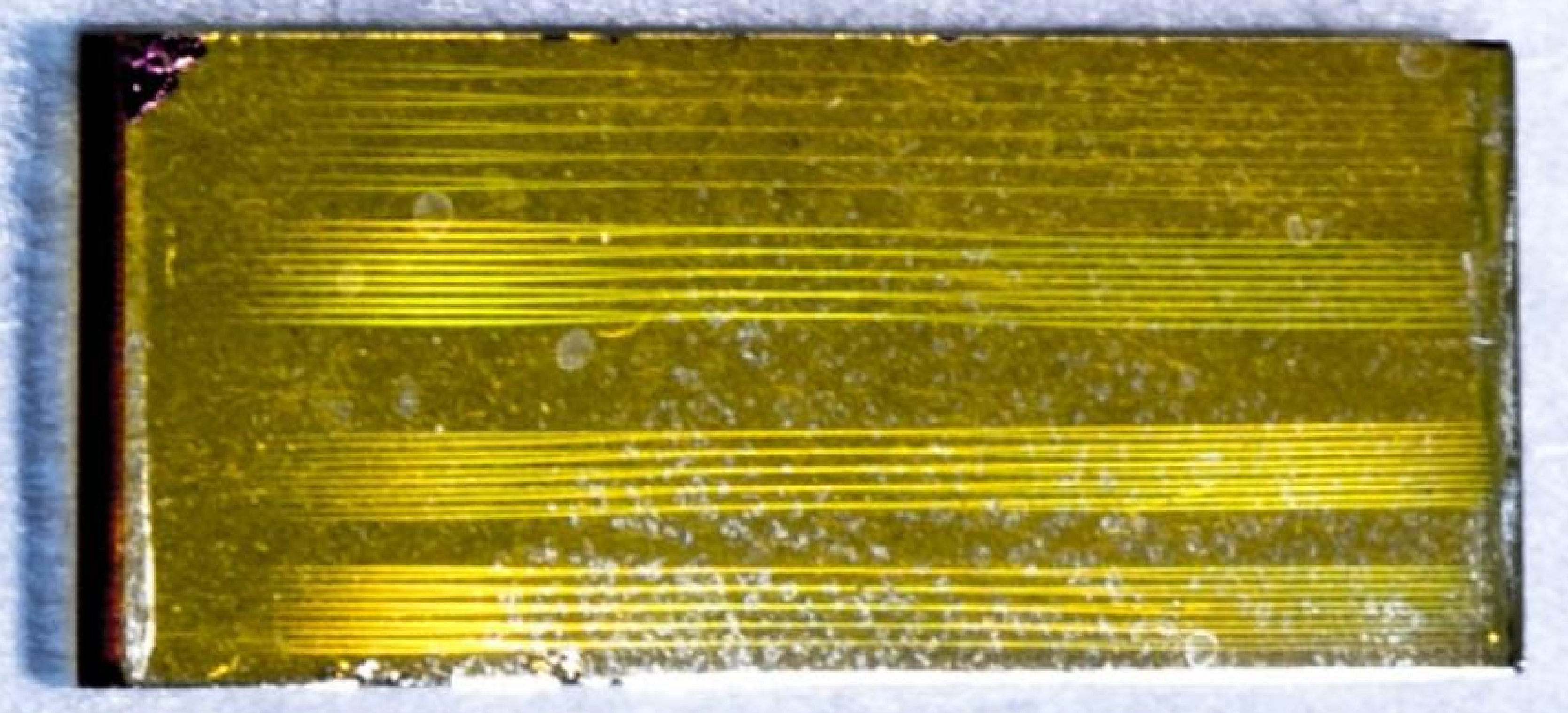}
    \hspace{2.5cm}
    \includegraphics[width=0.8\columnwidth]{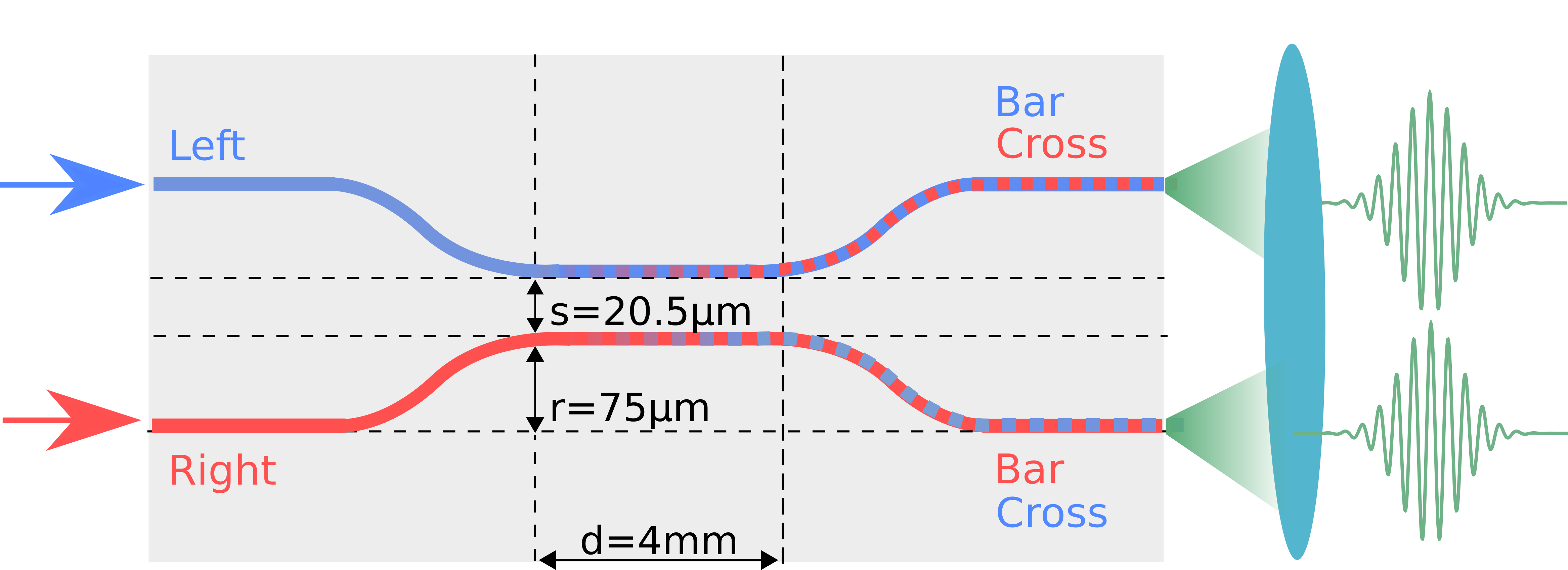}  
    \caption{Top: Integrated optics chip including 20 two-beam combiners written with different parameters. Bottom: Dimensions of the chosen two-beam combiner used in this paper and labeling of the waveguides. The terms \textit{bar} and \textit{cross} are used to distinguish the output of the initially excited waveguide from the evanescently coupled arm output.}
    \label{GLSchip}
\end{figure}
\subsection{Design and fabrication}\label{2.2}
The integrated optics chip presented in this paper (see Fig.\,\ref{GLSchip}\,top) is composed of commercial gallium lanthanum sulfide (GLS), a chalcogenide glass with refractive index $n = 2.3159 \pm 0.002$ at \SI{3.4}{\micro \meter}\footnote{Interpolated from refractive index measurements commissioned to VITRON GmbH, August 2016} and high transparency from 0.5 to \SI{9}{\micro \meter}, thus suitable for the astronomical L and M bands. In comparison to arsenic-based waveguides \citep{Vigreux2007}, GLS is a non-toxic material. 
The design of the device is a directional evanescent coupler (Fig.~\ref{GLSchip} bottom) as this represents a key building block for advanced functions such as ABCD phase and visibility estimators \citep{Colavita1999}. In the interaction area where the channel waveguides have the smallest separation, a fraction of the light injected in one arm is transferred into the nearby arm by evanescent coupling. A monochromatic splitting/coupling ratio of 50/50 can be obtained by optimizing the gap between the waveguides $s$ and the interaction length $d$ of the device. 
\newline
\indent The integrated optics chip was inscribed using the Jena laser writing facility composed of a femtosecond Yb:KGW laser at 1023\,nm launching 400\,fs pulses at 500\,kHz repetition rate, comparable to parameters used in \cite{Arriola2014}. 
 The waveguides are written using the multipath technique where cores are built up by a collection of 21 tracks, each spaced laterally by $\SI{300}{nm}$ with respect to the previous one.\newline
\indent Three design parameters were varied to estimate their effect on the performance of the device in the L band (cf. Fig.~\ref{GLSchip} bottom). Two values $r$ for the S-bend amplitude, namely 50 and \SI{75}{\micro \meter}, were considered to investigate the impact on the bending losses. The tested coupling lengths were 0\,mm and 4\,mm. The separation $s$ was varied between 20 and \SI{22}{\micro \meter} with \SI{0.5}{\micro \meter} steps as this parameter has a strong influence on the coupling as well. In total, twenty directional couplers were inscribed on the chip. The result is shown in Fig. 1 top where several of these couplers are visible on the chip with dimensions 25$\times$10$\times$1\,(mm)$^3$. Waveguides were written at a depth of \SI{200}{\micro \meter} from the top and bottom surface, respectively, and had measured cross sections of $ 7 \times \SI{25}{(\micro \meter)}^2$. 
We note that the impact of this design on the M-band performance was not considered at first in the selection of the sample that was eventually characterized.
\section{Laboratory setup}\label{labsetup}
The characterization setup, similar to that used in \cite{Labadie2007}, is based on a classical Michelson interferometer design (Fig.\,2). Two sources of light enable both broadband and monochromatic measurements: a CoolRed blackbody source (T=1500\,K) by Ocean Optics connected to a multimode infrared fiber from Thorlabs with \SI{400}{\micro \meter} core and a single-mode 5\,mW HeNe laser at \SI{3.39}{\micro \meter}. 
The sources are spatially filtered by 20 and \SI{25}{\micro\meter} pinholes, respectively, before being collimated with an f=50\,mm achromat and an f=150\,mm plano-convex lens, respectively. A pellicle beamsplitter (BS2) is used instead of a conventional thick beamsplitter to avoid differential dispersion in the interferogram. A Thorlabs Z812B delay line is used to adjust remotely the optical path difference in one arm by translating one flat mirror M1. Both mirrors can be adjusted in tip-tilt, and two images of the source can be created to be coupled to each input of the device by an f=50\,mm achromat. Using BS1, the laser beam can be independently aligned.
 \begin{figure}[t]
  \centering
    \includegraphics[width=0.95\columnwidth]{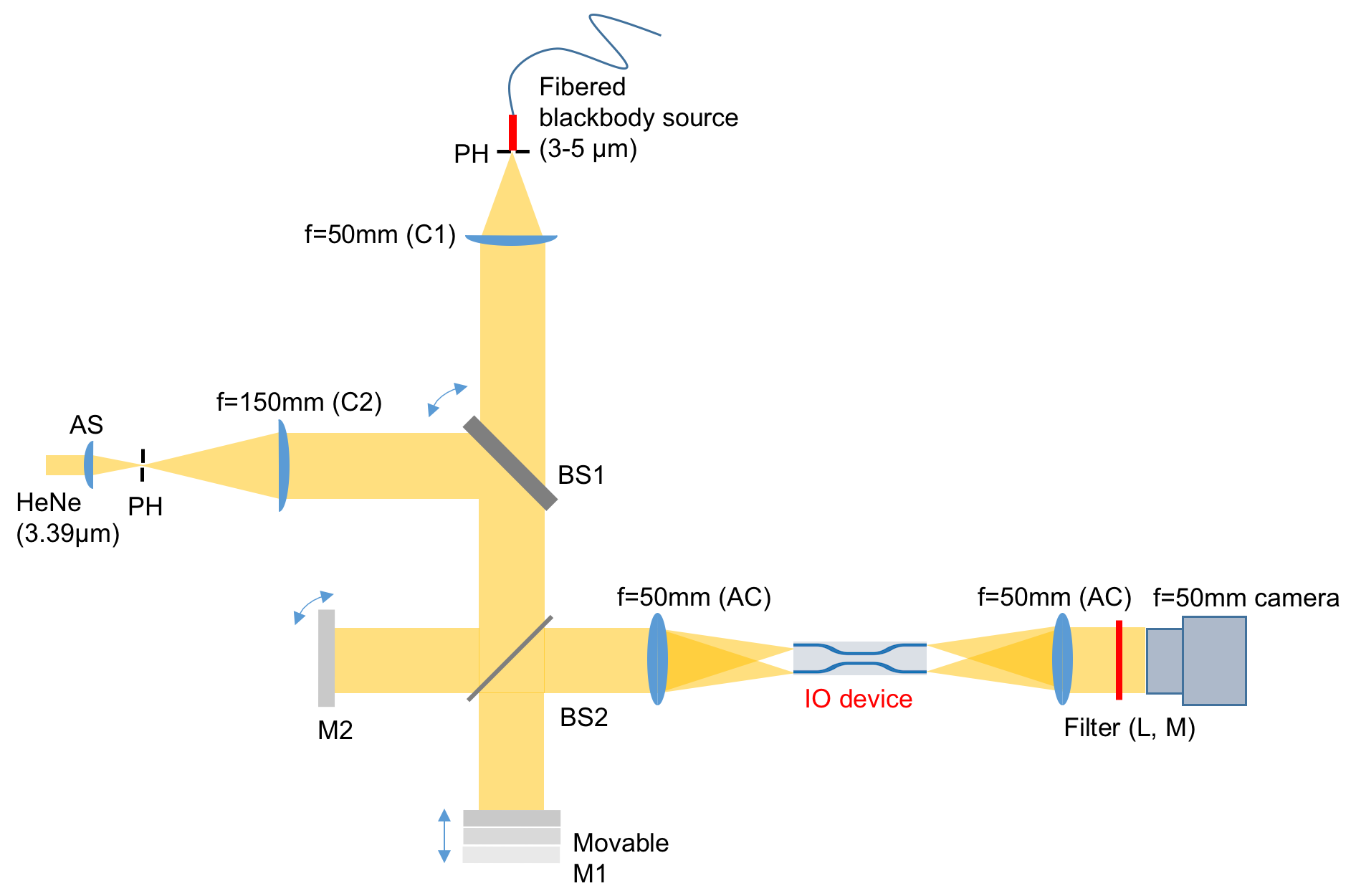}\label{setup}
    \caption{Layout of the experimental setup. AS: aspheric lens; PH: pinhole; C1, C2: collimator 1 \& 2; BS1: thick beamsplitter; BS2: pellicle beamsplitter; M1, M2: flat mirrors; AC: f=\SI{50}{mm} achromat; F: L or M broadband filter.}\label{bench}
\end{figure}
The two interferometric outputs are re-imaged by an f=50mm achromat and the 50\,mm camera objective onto the focal plane of the 5360S Infratech Camera. The IO chip can be fine-positioned in all three directions thanks to a high-precision XYZ translation stage.\newline
 \begin{figure}[b]
  \centering
    \includegraphics[width=1.0\columnwidth]{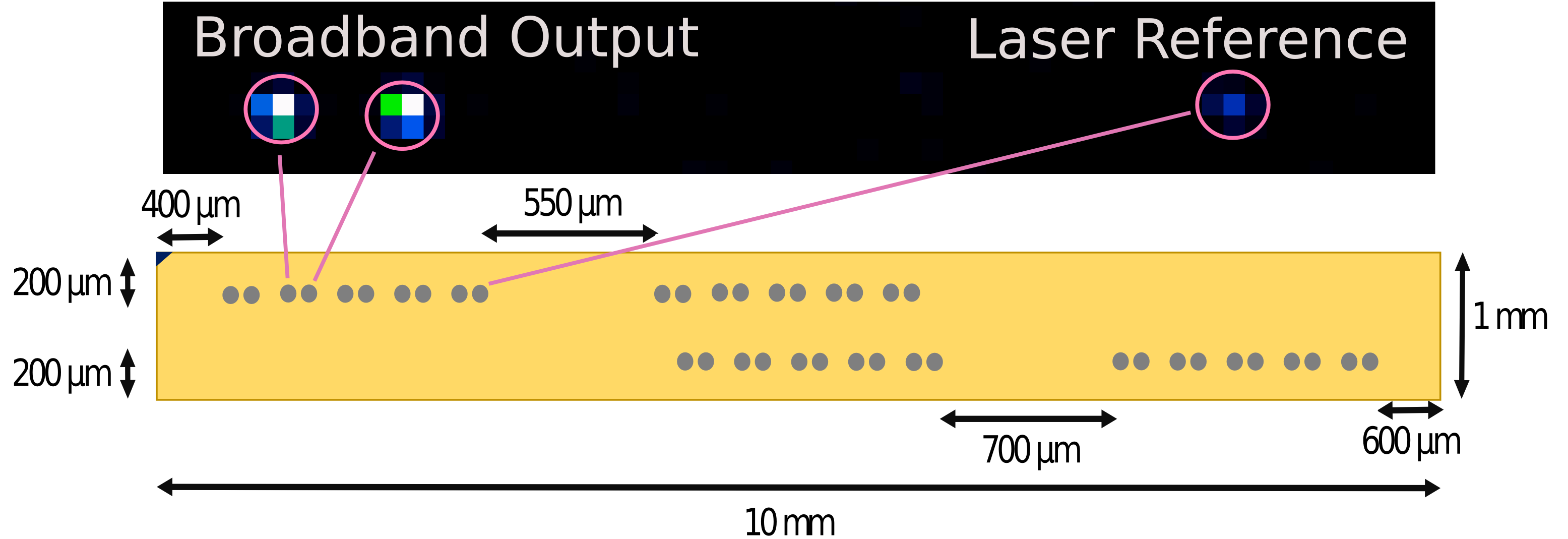}
    \caption{Outputs of the chip imaged with a magnification of 1 and focused on mainly one pixel. Next to the broadband output, the laser interferogram is recorded in a neighboring coupler. Below, the corresponding couplers from the chip (face-on view) are specified, in total 20 couplers. }\label{method}
\end{figure}
\indent A Michelson interferometer can operate as a Fourier transform spectrometer ,and we use this mode to measure the spectra of our device. A typical obstacle, however, is recording the true OPD as errors caused by the inaccurate and non-repeatable translation of the delay line directly translate into spectral errors in the Fourier space. We fix this problem by simultaneously recording the interferogram of the \SI{3.39}{\micro \meter} laser, see Fig.\,\ref{method}. The known fringe spacing is used to yield the true OPD and functions as a metrology channel \citep{Tepper2016}. 
\section{Results}
We present here	the	characterization of	the	properties of the directional coupler which are relevant for applications	in stellar interferometry. We report on the spectral splitting ratio, modal behavior, throughput, polarization properties, broadband interferometric contrast and chromatic dispersion. The impact of a cryogenic temperature cooling cycle on the performance of the component is presented as well.
\subsection{Splitting ratio over L and M band}
\begin{figure}[t]
   \centering
   \includegraphics{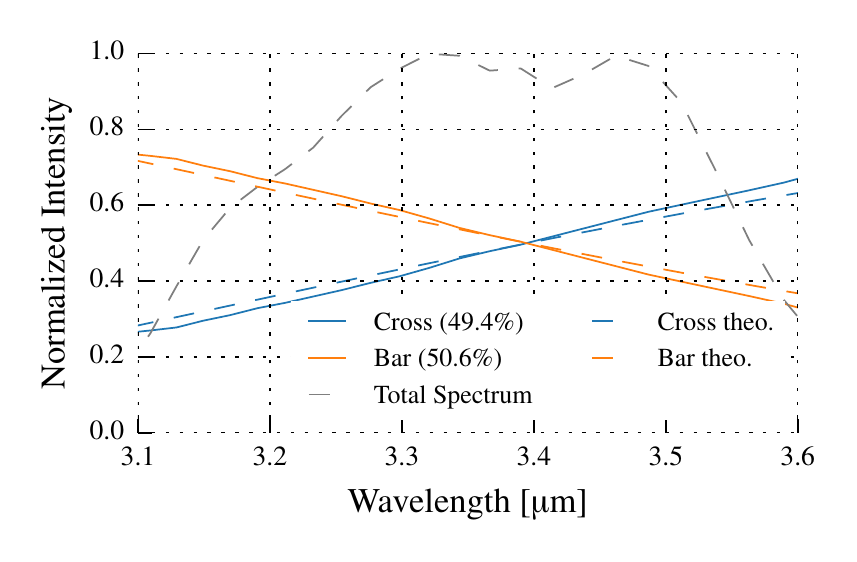} 
      \includegraphics{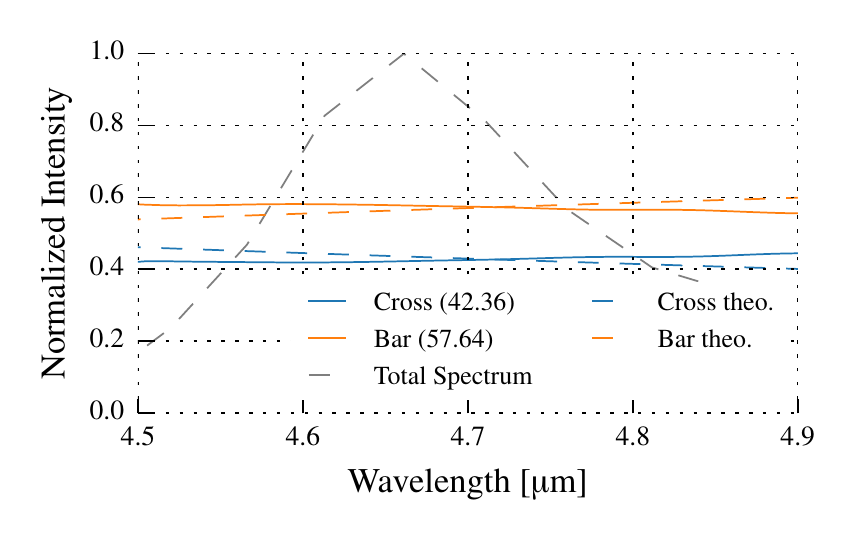} 
   \caption{Spectrally resolved coupling ratios for the left input over the L (top) and M (bottom) bands measured for the two outputs \textit{bar} and \textit{cross}. The numbers in the brackets refer to the spectrally integrated coupling. The dashed line shows the measured bandwidth of the experiment.}
   \label{LcouplingLeft}%
   \end{figure}

We were provided with 20 couplers written with different parameters in one single chip as seen in Fig.\,\ref{GLSchip} top and \ref{method}. In order to find the most suitable coupler for L-band interferometry, we first investigated the broadband splitting ratio defined as $P_{cross}/(P_{cross}+P_{bar})$, where $P$ is the power in the respective channel (Fig.\,\ref{GLSchip} bottom). The splitting ratios, which depend on the interaction length and the separation of the waveguides in the interaction area, are found to range from 10 to 85\% for the different couplers of the chip. The best coupler, i.e., closest to 50/50, shows a splitting ratio of 49.4\%. The design properties of this particular coupler are $d$=4\,mm and $s$=\SI{20.5}{\micro \meter} (see Fig.\,\ref{GLSchip} bottom).\newline
\indent Using Fourier transform spectroscopy, we measured the chromaticity of the splitting ratio. By injecting the two input beams in the same input and varying the OPD, we yield one interferogram for each output from which the respective spectrum is derived (see Fig.~\ref{LcouplingLeft} top). 
We find a linear trend of the splitting ratio ranging from about 30\% at \SI{3.1}{\micro \meter} to 70\% at \SI{3.6}{\micro \meter}. Also shown as a dashed line is the normalized spectrum of the totally transmitted flux (i.e. the sum of the two output fluxes), which is the product of the transmission curves of the respective broadband filter, of the optical bench and of the waveguide. The chromatic splitting ratio for the right input was also measured and shows the same behavior.\newline
\indent
Using the same coupler in the astronomical M band, we found an imbalance in the broadband splitting ratio of 42.36\%. The spectral dependence, however, is found to be flatter (see Fig. \ref{LcouplingLeft} bottom). This coupler is used for all further measurements, except for the spectral transmission (Fig. \ref{transmission}, channel waveguide instead) and the temperature test in Sect. \ref{temp}.\newline
\indent Theoretically, the splitting ratio for the directional coupler can be derived as described in \cite{Snyder1983}. As the calculation is sensitive to the cross section of the waveguide, which is not known precisely, we used a simplified, weakly wavelength-dependent approach for the coupled power, sin$^2$($K\cdot d/\lambda$), $K$ being a real scalar constant and $d$ the interaction length. Only a discrete set of $K$ values correspond to a 50/50 splitting at a given wavelength, which is here taken at \SI{3.4}{\micro \meter} and \SI{4.7}{\micro \meter}, respectively, for the L and M band. From these values, the correct $K$ can be univocally determined by comparing the slope to the experimental data (Fig.\,\ref{LcouplingLeft}). The $K$ values ($K_L=2.0 \cdot 10^{-3}$ and $K_M=0.84 \cdot 10^{-3}$) indicate that we operate close to 0.75 and 0.25 beat length, respectively, at the above-mentioned reference wavelengths for the L and M band.

\subsection{Modal behavior}
\begin{figure}
\centering
\raisebox{-0.2cm}[0cm][0cm]{
  \includegraphics[width=4.1cm]{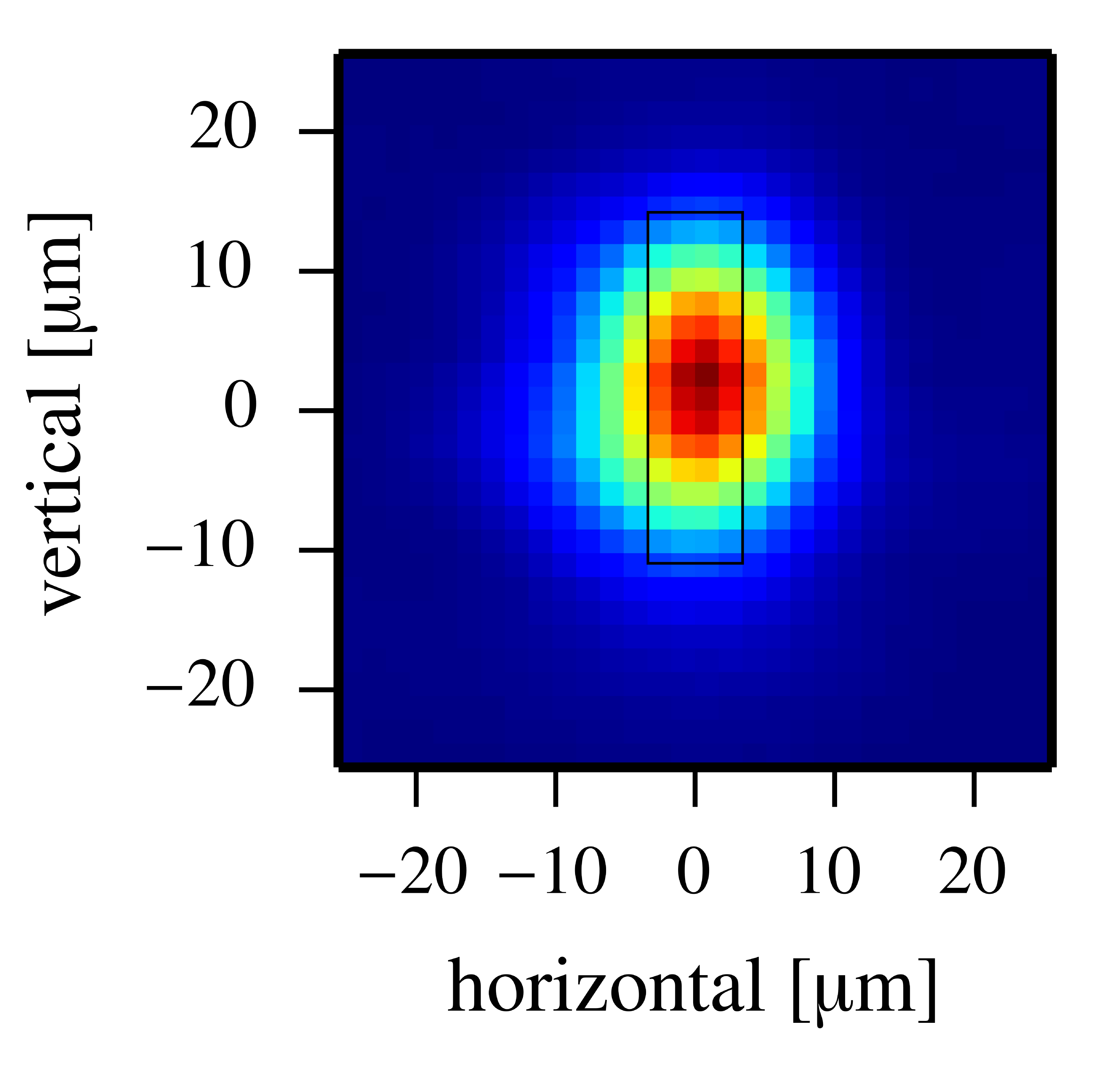}
  }
  \includegraphics[width=3.9cm]{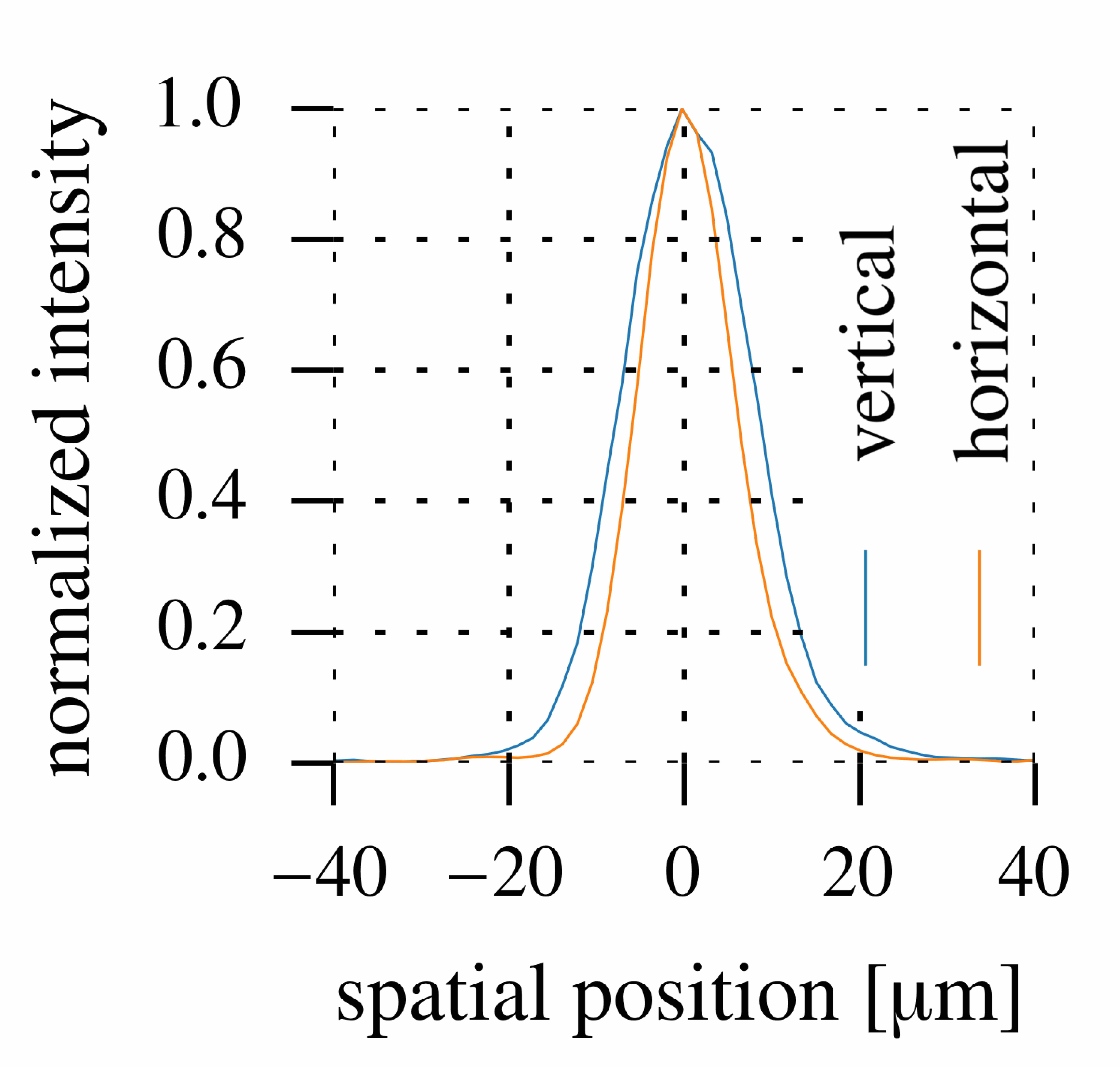}
  \caption{Left: Left output of the two-telescope combiner with a magnification of 17 at \SI{3.39}{\micro \meter}. The black rectangle illustrates the size of the waveguide cross section. Right: Respective vertical and horizontal Gaussian-like mode profiles before deconvolution.}
  \label{mode}
\end{figure}

For the purpose of wavefront filtering \citep{Ruilier2001}, it is essential that the waveguides exhibit single-mode behavior over the considered bandwidths. Figure \ref{mode} shows an image of one of the two outputs and the respective mode profile. We use an optical system with a working f-number of f/36  to magnify and image the MFD onto our sensor with a pixel size of \SI{30}{\micro \meter}. We find a Gaussian-like shape as expected for single-mode waveguides, with 1/e$^2$ measured mode field diameters (MFD) of $28.2 \pm 0.3$\,\SI{}{\micro \meter} (vertically) and $(22.6 \pm 0.2)$\,\SI{}{\micro \meter} (horizontally) at \SI{3.39}{\micro \meter}. The measured FWHM point spread function (PSF) of the optical system is \SI{8.4}{\micro \meter} (\SI{8.5}{\micro \meter} theoretically). 
We derive the true MFD as the diameter of the Gaussian waveguide mode, whose convolution (i.e., the convolution of the Gaussian mode) with the PSF results in the measured MFD.
	 We obtain \SI{16.3}{\micro \meter} and \SI{24.8}{\micro \meter} for the horizontal and vertical direction, respectively. This can be related to the waveguide cross section to obtain the refractive index contrast. As an approximation, we used a rectangular waveguide model with a step-index\footnote{Calculated using the online tool http://www.computational-photonics.eu/eims.html}. Indeed, the cross section is only known with an accuracy of $\pm$\,\SI{2}{\micro \meter}, and the true index profile is unknown. Therefore, we can only give a range for the refractive index contrast from $3 \cdot 10^{-3}$ to $4 \cdot 10^{-3}$.
\newline
\indent Single-mode behavior is classically tested against the presence of a second higher order mode by changing the input coupling conditions, such as inserting a small lateral displacement of the injection spot \citep{Ho2006}. We applied such a displacement in the vertical and horizontal direction and we did not find any deviation from the near-Gaussian shape. 
\subsection{Throughput}
	In this section, we present the total and spectrally resolved relative throughputs of the coupler. \newline
	\indent The total throughput in the L band is estimated by injecting into one of the two inputs, and measuring the sum of the flux from the two outputs. Dividing this number by the input beam flux (measured by removing the sample in front of the imaging system), we obtain the throughput of the component. 
        For the throughput experiment we used input beam diameters of 8, 11, 14, and \SI{16}{\milli\meter} in order to find the optimal input coupling efficiency. Table \ref{throughput} shows the throughputs for the different injection beam diameters. We find the optimal throughput of 25.4\% at \SI{11}{\milli\meter}. This corresponds to a numerical aperture of 0.11 and a $\Delta n=(3.1 \pm 0.1) \cdot 10^{-3}$, which is in line with the change in refractive index derived in the previous section. 
        \newline
        \indent The total throughput can be written as $T=(1-R_F)^2 \cdot C \cdot PL \cdot B^2$, where $C$ is the input coupling efficiency and $R_F$ the Fresnel reflections at the input and output facets. The values of $P$ and $B$ account for the reduced transmission due to propagation losses per cm and bending losses per bend, respectively. The quantity $L=\SI{2.5}{cm}$ is the length of the component. 
        For the coupling efficiency we assume the optimal coupling between a Gaussian and an Airy pattern given by $C$=81\% \citep{Toyoshima2006}. 
         Due to the high refractive index $n=2.316$, a Fresnel reflection coefficient of $R_F=15.7\%$ per facet is found. This can be mitigated by an AR coating, which would raise the throughput to 36\%. Bending losses were separately measured to (0.6$\pm$0.2)\,dB/bend. Using these numbers and taking the averaged throughput for the \SI{11}{\milli\meter} beam diameter, we estimate $P$ to be $\SI{0.94}{dB/cm} \pm \SI{0.29}{dB/cm}$. We stress that this number is a rough approximation, due to the large error bar on the bending losses, and is also very sensitive to the input coupling efficiency. For instance, for a coupling efficiency of 70\%, the value $P$=0.69\,dB/cm would be derived from our measurements. 
    \begin{table}
\caption{Throughput of the \SI{25}{\milli\meter} two-telescope combiner in L band for different input beam diameters. Absolute accuracy on the order of 2\% due to error propagation of imperfect adjusting of the iris.}             
\label{throughput}      
\centering                          
\begin{tabular}{c c c c c}        
\hline\hline                 
Beam diameter & \SI{8}{\milli\meter} &\SI{11}{\milli\meter} &\SI{14}{\milli\meter}& \SI{16}{\milli\meter}\\  
\hline                        
Left Input & 20.4\% & 22.5\% & 16.6\%   & 13.8\% \\
Right Input  & 24.5\% & 28.4\% &   24.1\% & 21.9\%           \\
Average  & 22.4\% & 25.4\% &   20.4\% & 17.8\%  \\
\hline                                   
\end{tabular}
\end{table}
    \begin{figure}[h]
   \centering
   \includegraphics[width=\hsize]{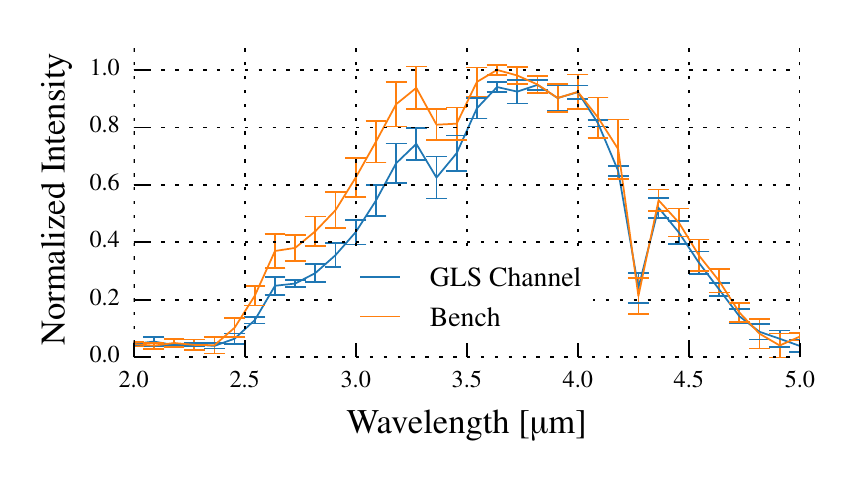}
   \caption{Normalized transmission of the optical bench and transmission of the optical bench including a GLS channel waveguide. The GLS data is upscaled without exceeding the bench data. For each case, three datasets were taken and the error bars show the standard deviation.}
              \label{transmission}%
    \end{figure}
    \newline
    \indent The transmission spectrum of the component was measured by Fourier transform spectroscopy. The two beams are injected into the same channel waveguide and by scanning the OPD, an interferogram is recorded from which the relative spectrum is derived. This is directly compared to the normalized transmission of the bench (see Fig. \ref{transmission} for details). 
    We find a very good match between the two spectra, which indicates a flat spectral response of the waveguide. The location of the CO$_2$ dip at $\approx$ \SI{4.26}{\micro\meter} is measured at the expected position in both transmission spectra and confirms the validity of the OPD correction method. 
\subsection{Polarization properties}\label{pol}
Instrumental contrasts measured with a long-baseline interferometer are highly sensitive to polarization mismatches resulting from differential stress and birefringence between the two arms \citep{Berger1999}. When operating with unpolarized stellar light, it is important that the differential birefringence is minimized so that the polarization alteration that may arise from the IO component is similar in each arm, hence reducing the visibility loss effects.\newline
\indent We investigate the polarization properties of the coupler at \SI{3.39}{\micro\meter}. We look at one output and test the difference in polarization for the two arms by exciting the left and the right input one at a time. By placing a half-wave plate before the component and an analyzer behind the output collimation lens, we analyze the change in polarization state for different incoming polarization angles. Two different quantities are measured: the change in polarization angle and to what extent linear polarization is transformed into elliptical polarization. The latter is quantified as the polarization contrast $P_c=\frac{I_{max}-I_{min}}{I_{max}+I_{min}}$, where $I_{max}$ and $I_{min}$ are the maximum and minimum intensities, respectively, found by turning the polarizer. Consequently, $P_c=1$ for a linear polarization and $P_c=0$ for a circular polarization.\newline
\begin{figure}[!h]
    \centering
   \includegraphics{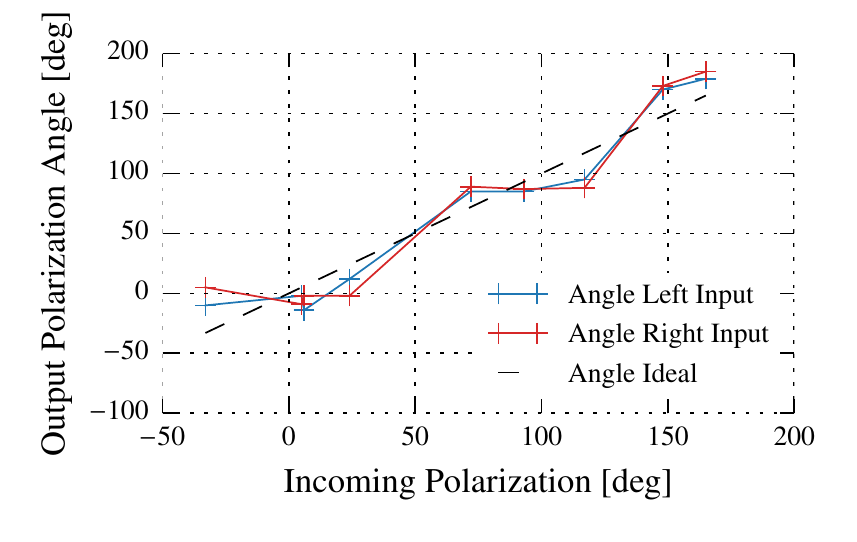}
   \caption{Polarization angle of the output ellipse (cf. Fig.\ref{contrasts}) for the left output when injecting into the left and right input, respectively. For the input, we used the linearly polarized HeNe laser at \SI{3.39}{\micro \meter} at different polarization angles (depicted on the x-axis). The dashed line depicts an unaltered polarization angle. Qualitatively, there is very little or no difference between the two output polarization angles. }
              \label{angles}%
   \end{figure}
   \begin{figure}[!h]
   	\centering
   	\includegraphics{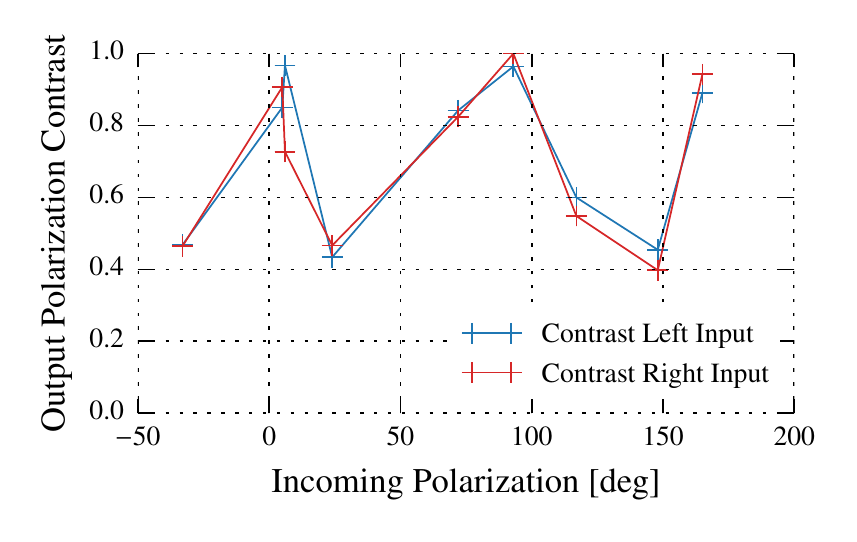}
   	\caption{Change in polarization contrast ("ellipticity") for the left output when injecting into the left and right input, respectively, for incoming linear polarization of different angles.As for the angle of the ellipse (cf. Fig.\ref{angles}), there is very little or no difference in the ellipticity between the two outputs.}
   	\label{contrasts}%
   \end{figure}
\indent From Fig. \ref{angles}, we find that the polarization angle is similarly well maintained for both inputs to within 20$^\circ$. Furthermore, for input angles close to 0, 90, and 180$^\circ$, the polarization state is almost unchanged and remains linear, as seen in Fig. \ref{contrasts}. For other angles, the linear polarization is transformed into an elliptical polarization with the contrast dropping down to 0.4. The same result is found by measuring the right output. The important result from Figs. 7 and 8 is that, for a given output, the change in polarization is almost identical for both arms, making differential  polarization effects small. The polarization properties of the coupler can be explained by the rectangular waveguide cross section, which induces shape birefrigence \citep{Marcuse1974}. 
\subsection{Interferometric contrast and differential dispersion}
In the following sections interferograms over different bandwidths using the evanescent 2$\times$2 coupler are presented. The interferograms are recorded for linearly polarized laser light and unpolarized broadband light. The recording and post-processing is as follows. First, the photometric channels are recorded. Then, the interferogram is recorded. The laser signal is recorded simultaneously to yield the true OPD as described in Sect. \ref{labsetup}.
\subsubsection{Monochromatic interferogram at \SI{3.39}{\micro\meter}}
\begin{figure}
   \includegraphics{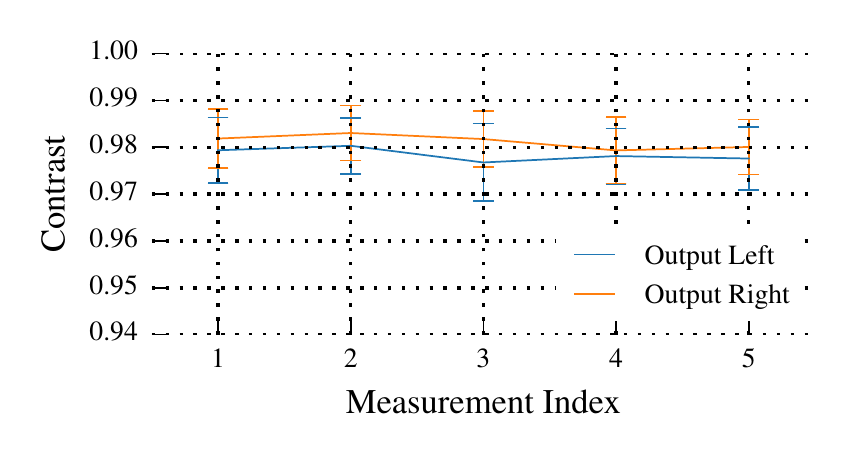}
   \caption{Monochromatic interferometric contrast of the two outputs at \SI{3.39}{\micro\meter} for a series of measurements. The error bars show the standard deviation in contrast for the respective measurement.}
              \label{mono}%
    \end{figure}

First, using the vertically linearly polarized HeNe laser at \SI{3.39}{\micro\meter}, a monochromatic interferogram is recorded. Each interferogram is scanned over approximately 125 fringes. From the pairs of local maximums and minimums, 125 contrast values can be calculated. The average contrast and its standard deviation are shown in Fig.\,\ref{mono}. This measurement was repeated five times and averaged contrasts of 97.8$\pm$0.6\% and 98.1$\pm$0.6\% for the two outputs are found (sampling effects are negligible). Repeated testing showed that the measured contrast does not change for an incoming polarization angle of 45$^\circ$. This is in line with Figs.\ref{angles} and \ref{contrasts}, which show that differential birefrigence between the two arms is small for any incoming angle.\newline
\indent In comparison, for an interferogram recorded through a single-channel waveguide, where no differential birefrigence is present, we find 99.4\% with a standard deviation of 0.2\%. This loss in contrast of $\sim$1.4\% would correspond to a mismatch in polarization angle of 0.2\,rad.
\subsubsection{L band interferogram}
\begin{figure}[h!]
   \centering
   \includegraphics{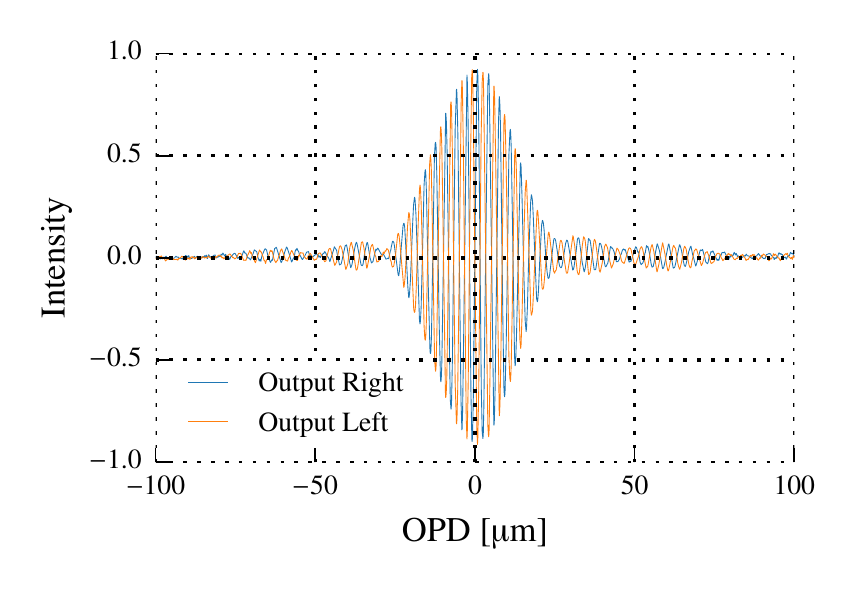}
   \caption{Experimental L-band interferogram of the two interferometric outputs of the coupler after photometric correction. A broadband contrast of 94.9\% is measured. The spectral shape of the bandwidth is shown in Fig.\,\ref{LPhase}. The respective interferograms of the two outputs are shifted by half a wavelength, i.e., $\pi$-phase shifted.}\label{LBandInterferogram}%
    \end{figure}

The L band filter, covering the range from 3.1 to \SI{3.6}{\micro \meter}, as seen in Fig.\,\ref{LcouplingLeft} top, is inserted into the collimated beam. Figure \ref{LBandInterferogram} shows the two recorded interferometric outputs. We find a high contrast of 94.9\%. In addition to the high broadband contrasts, the $\pi$-phase shift between the two outputs resulting from energy conservation is observed with excellent repeatability over the coherence length.
     \begin{figure}[h!]
       \centering
       \includegraphics{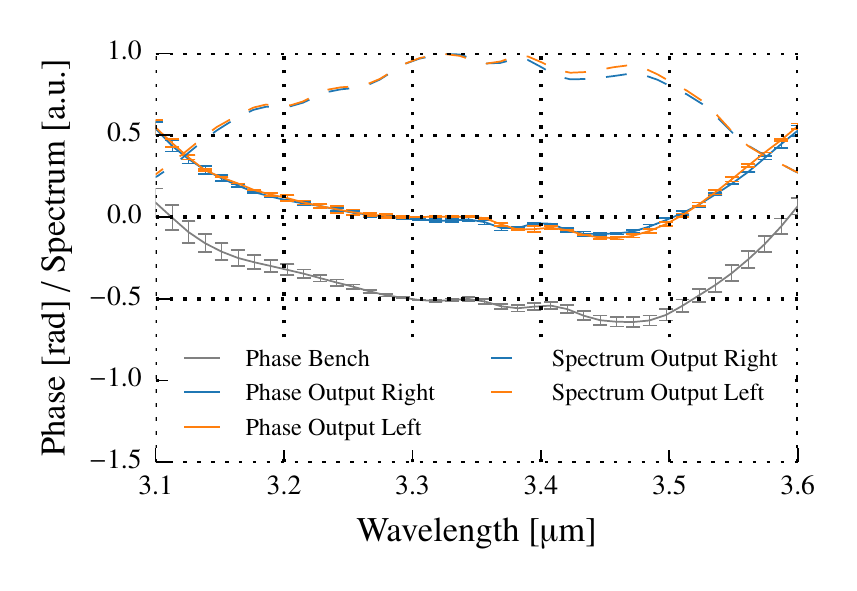}
       \caption{Phases for the two chip outputs as well as the phase of experimental setup without chip (lowered by 0.5\,rad for visualization) over the L band. The value $\pi$ was subtracted from one of the chip outputs to demonstrate the phase shift. The respective spectra are shown as dashed lines.}
                  \label{LPhase}%
        \end{figure}
   \newline
   \indent
   In order to estimate the level of differential dispersion, we calculated the phase of the interferogram through the real and imaginary part of its Fourier transform. After removing the linear part in wavenumber $2\pi x_0 \sigma=2\pi x_0 /  \lambda$, which relates to the zero-OPD position $x_0$  \citep{Foresto1995}, the non-linear term remains, which accounts for the overall differential dispersion. In order to disentangle any dispersion that may arise from the experimental setup, we also measured the phase for an interferogram without the IO chip. The results are shown in Fig.\,\ref{LPhase}. We find that the phase variation across the band is mainly determined by the phase from the testbench itself and that the combiner is close to dispersion free with a standard deviation of 0.04\,rad across the band. For visualization purposes $\pi$ was subtracted from one of the two chip outputs. The excellent overlap of the two phase curves reflects the clean $\pi$-phase shift visible in the interferogram. \newline
\indent
The present dispersion can be further quantified through the dispersion parameter defined as $D:=\frac{d(\tau_g)}{d\lambda}$, i.e., the derivative of the group delay with respect to wavelength. From \cite{Foresto1995}, the (differential) dispersion parameter can be related to the phase curvature, i.e., the second derivative of the phase with respect to the wavenumber, through
\begin{equation}
\frac{d^2 \Phi}{d \sigma^2}=-2\pi c \lambda^2 \cdot (L \Delta D +   D \Delta L).
\end{equation}
 Assuming that the two channels have identical lengths $\Delta L=0$, the quantity of interest is the difference in dispersion parameters of the two channels multiplied by the length of the component $L \cdot \Delta D=L \cdot (D_2-D_1)$. Before calculating the dispersion parameter, we subtracted the phase of the bench from the phase of the beam combiner interferogram. Then, calculating the second derivative of the phase averaged over the bandwidth, we find $L\,\Delta D=3.6\cdot 10^{-5}$\,ps/nm. In comparison, \cite{Foresto1995} finds $1.8\cdot 10^{-4}$\,{ps/nm} for a moderately dispersed interferogram. For our case, we then find a differential dispersion parameter of $\Delta D = 1.4$\,ps/(km$ \cdot$ nm) with a standard deviation of 15.7\,ps/(km$ \cdot$ nm) across the band. Both values must be read as an upper limit as the phase variation due to the combiner is within the error bars (see Fig.\,\ref{LPhase}).
\subsubsection{M Band interferogram}
\begin{figure}[h!]
   \centering
   \includegraphics{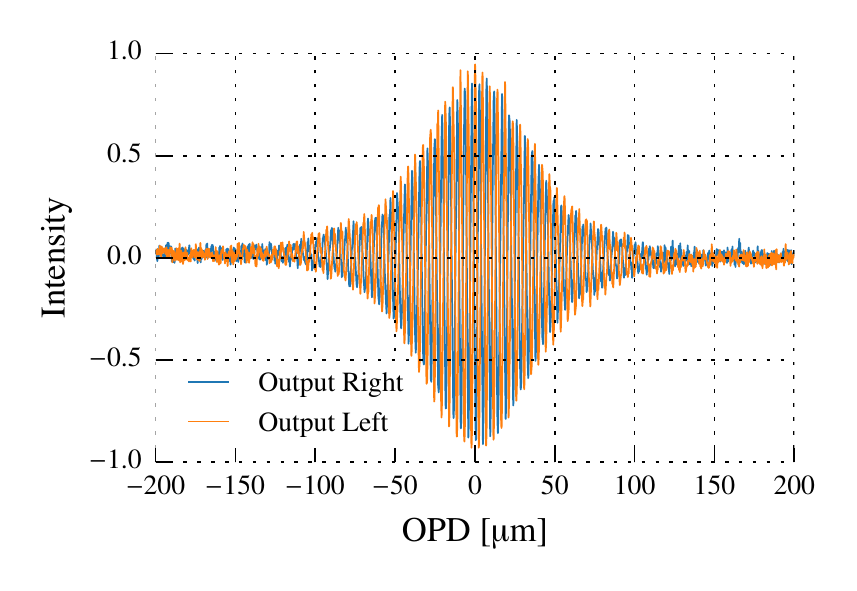}
   \caption{Experimental M-band interferogram of the two interferometric outputs of the coupler after photometric correction. A broadband contrast of 92.1\% is measured. The spectral shape of the bandwidth is shown in Fig.\,\ref{MPhase}. The respective interferograms are not perfectly $\pi$-phase shifted, as can be seen from Fig.\,\ref{MPhase}.}
              \label{MBandInterferogram}%
    \end{figure}

\begin{figure}
      \centering
      \includegraphics{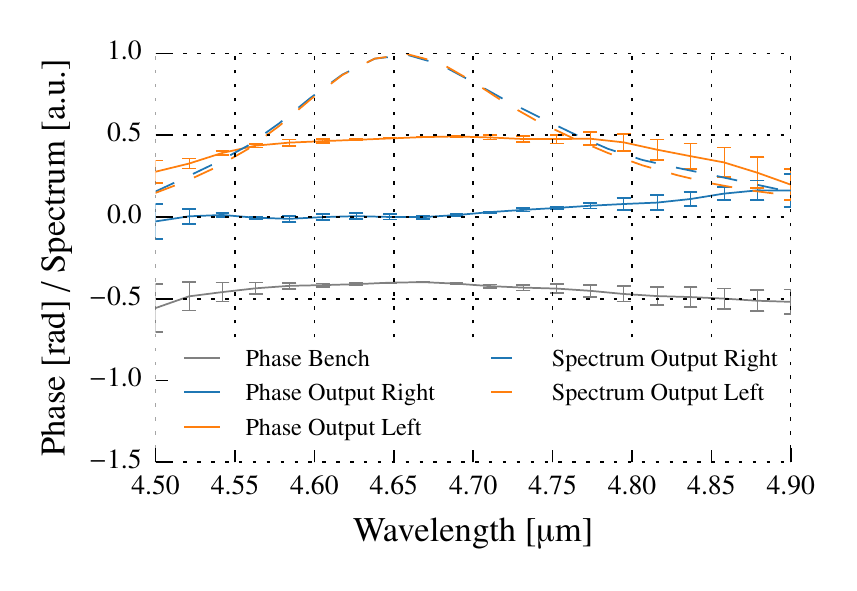}
      \caption{Phases for the two chip outputs and the phase of experimental setup without chip (lowered by 0.5\,rad for visualization). After subtraction of $\pi$ a residual phase difference of 0.4\,rad between the two outputs remains. The respective spectra are shown as dashed lines.}
                 \label{MPhase}%
       \end{figure}

We measure the M-band interferogram through the same component and find a contrast of 92.1\% (see Fig.\,\ref{MBandInterferogram}). By applying the same procedures as for the L-band interferogram, we obtain the phase and the differential dispersion parameter. Here we find a relatively flat phase both for the bench and for the bench including the combiner (see Fig.\,\ref{MPhase}). After subtracting the bench phase, we are left with a standard deviation of 0.07\,rad across the band. The differential dispersion parameter of the combiner $\Delta D = 2.8$\,ps/(km$\cdot$nm) is about twice as large as for the L band with a standard deviation of 17.9\,ps/(km$\cdot$nm) across the band. This may be due to the larger deviations towards the edge of the spectrum which result from the lower flux in the M band, as can be seen from the increasing error bars. After removal of $\pi$ between the two outputs of the chip, we still find an offset of about 0.4\,rad between the two phases. This shows that the phase shift slightly deviates from $\pi$, although this cannot be easily seen in Fig.\,\ref{MBandInterferogram}.%


   \subsection{Temperature test}\label{temp}
Another similar spare coupler was cooled down to \SI{120}{K} in a cryostat over a approximately ten hours. After bringing the sample back to room temperature, no physical changes, e.g. cracks in the glass, were observed. The coupler was interferometrically characterized in the L band before and after the cooling down. In this one-time test, no major difference in throughput or interferometric properties was found. Thermal shocks, however, caused by dropping the sample directly into liquid nitrogen at \SI{77}{K}, lead to cracks in one of two instances. As cryogenic conditions are typically used for mid-IR instruments, further testing is required to understand the optimal cooling rate $\Delta T / \Delta t$ for our component.
   \begin{table}[h!]
\caption{Overview of the coupler properties. Numbers are averaged over the two inputs for the splitting and averaged over the two outputs for the phase and contrast.} 
\label{contrast}      
\centering                          
\begin{tabular}{c c c c}        
\hline\hline                 
     &  \SI{3.39}{\micro\meter} & L band & M band \\  
\hline                        
            integrated splitting & - & 49.4\% & 42.3\%      \\
            spectral splitting variation  & - & 35.5\% & 5.1\%             \\
            throughput & - & 25.4\% & - \\
            diff. polarization & 0.2\,rad& & \\
                               & \& see Sect.\ref{pol}& & \\
phase variation & - & 0.04\,rad & 0.07\,rad \\
             contrast & 98.0\% & 94.9\% & 92.1\% \\
\hline                                   
\end{tabular}
\end{table}
 \section{Discussion and conclusion}
We have presented a full characterization of a laser-written 2$\times$2 integrated-optics beam combiner in the 3\,-\,\SI{5}{\micro \meter} mid-infrared range. The measured properties are summarized in Table 2.\newline
\indent From twenty different test couplers, we chose to characterize the component with a spectrally integrated splitting ratio close to 50/50 over the L band. The measured splitting ratio shows a maximum imbalance of 30/70 across a relatively large bandwidth of $\Delta \lambda$=\SI{0.5}{\micro \meter} with the balanced splitting being at \SI{3.4}{\micro \meter}. The chromatic dependence is larger than has been measured for the GRAVITY beam combiner \citep{Benisty2009}. However, the latter was measured over a narrower bandwidth $\Delta \lambda / \lambda= \SI{0.12}{\micro \meter} / \SI{1.6}{\micro \meter} =0.08$, whereas our result covers a broader bandwidth with $\Delta \lambda / \lambda=\SI{0.5}{\micro \meter} / \SI{3.4}{\micro \meter} =0.15$. Also, GRAVITY used 2$\times$2 asymmetric tapered couplers which allowed for an achromatic design. As we operate at a beat length of 0.75 in the L band, one immediate step towards a more achromatic design will be to test shorter interaction lengths to come closer to an achromatic coupler. In the M band, the maximum imbalance is found to be 40/60 over a bandwidth of $\Delta \lambda$=\SI{0.4}{\micro \meter}. These are encouraging results given that a directional coupler is by definition a chromatic device. Further flatness of the spectral splitting ratio can be obtained with asymmetric coupler design \citep{Takagi1992} or other broadband design, as in \cite{Hsiao2010}.\newline
\indent We find that the coupler exhibits some birefringence which supports non-degenerate quasi-TM and quasi-TE polarization modes. However, because of the small differential polarization effect, the measured interferometric contrast in broadband unpolarized light remains high. Contrasts were measured of 98\% at \SI{3.39}{\micro \meter} as well as 94.9\% over the L band and 92.1\% over the M band. 

\indent Finally, we quantitatively  assessed the dispersion parameter. Considering the length of the component, the estimated dispersion parameters $D$ of 1.4 and 2.8\,ps/(km\,$\cdot$\,nm), respectively, have little impact on the broadband interferograms. The deviation from 100\% in contrast in L band can be roughly attributed to several effects: about 2\% due to the chromatic splitting, about 1.4\% due to differential birefrigence and the rest possibly due to the dispersion of the experimental setup. In the M band this assignment is more difficult as the signal-to-noise ratio is much lower in that case.
\newline
\indent Our single-mode coupler shows a total throughput of 25.4\%, including Fresnel, coupling, bend and propagation losses. Since a meaningful requirement for astronomical applications is the total throughput (after mitigation of the Fresnel losses), a potential objective is to revise the design of the device for a better trade-off between the propagation losses (higher for a longer component) and the bending losses (higher for a shorter component). Given that propagation losses can be as low as 0.3\,dB/cm in similarly laser-written channel waveguides (Thomson 2016, personal communication), we are optimistic that by further optimization of the writing parameters, the total throughput can be significantly increased. We found a refractive index modification of about $\Delta n=3 \cdot 10^{-3}$, which is about five times greater than the negative laser writing in ZBLAN reported in \cite{Gross2015}. A large $\Delta n$ is crucial in reducing bending losses and essential in order to implement more complex optical designs without greatly increasing the overall losses.\newline
\indent In the view of a multi-aperture (4+) beam combiner, more advanced optical functions are targeted in the next phases of our technology roadmap. Also, the development of a fiber-fed system and the interface between the infrared fibers and the chip will be addressed.
\begin{acknowledgements}
LL, RD, SM, J-UP, RT, and JT acknowledge the financial support from the German Ministry of Education and Research (BMBF). Special thanks go to Michael Wiest who helped carrying out the cryogenic measurements. \newline
The authors thank the anonymous referee for the constructive remarks which helped to improve the quality of the paper.
\end{acknowledgements}
\bibliographystyle{aa}
\bibliography{report.bib}
\end{document}